%% file: levenson.tex
\documentclass[12pt, preprint]{aastex}
\newcommand\ea{et al.\ }
\newcommand\rosat{{\it ROSAT}}
\def\chandra{{\it Chandra}}

\def\o3{[\ion{O}{3}]}
\def\s2{[\ion{S}{2}]}
\newcommand\kms{\ifmmode{\rm\,km\,s^{-1}}\else{${\rm\,km\,s^{-1}}$}\fi}
\newcommand\psc{\ifmmode{\rm\,cm^{-2}}\else{${\rm\,cm^{-2}}$}\fi}
\shorttitle{Shell and Cloud Shocks in the Cygnus Loop}
\shortauthors{Levenson et al.}
\slugcomment{Accepted for publication in the Astrophysical Journal}

\begin{document}
\title{Shell Shock and Cloud Shock: Results from 
Spatially-Resolved X-ray Spectroscopy with {\it Chandra} in the Cygnus Loop}

\author{N. A. Levenson}
\affil{Department of Physics and Astronomy, Bloomberg Center, Johns Hopkins University, Baltimore, MD 21218, and
 Department of Physics and Astronomy, University of Kentucky, Lexington, KY 40506}
\email{levenson@pa.uky.edu}
\and
\author{James R. Graham and Julie L. Walters}
\affil{Department of Astronomy, University of California, Berkeley, CA 94720}
\email{jrg@astron.berkeley.edu, jwalters@astron.berkeley.edu}

\begin{abstract}
We use the {\it Chandra X-ray Observatory} to
analyze interactions of the blast wave and the 
inhomogeneous interstellar medium on the western
limb of the Cygnus Loop supernova remnant.
This field of view includes 
an initial interaction between the blast wave and
a large cloud, as well as the encounter of the shock front and the
shell that surrounds the cavity 
of the supernova progenitor. 
Uniquely, the X-rays directly trace the 
shock front in the dense cloud, where we measure temperature
$kT = 0.03$ keV.
We find $kT \approx 0.2$ keV in regions where 
reflected shocks further heat previously-shocked material.
Applying one-dimensional models to these interactions,
we determine  
the original blast wave velocity 
$v_{bw} \approx 330 {\rm \,km\, s^{-1}}$
 in the ambient medium.
We do not detect strong evidence for instabilities
or non-equilibrium conditions on the arcsecond scales we resolve. 
These sensitive, high-resolution data
indicate no exceptional abundance variations in
this region of the Cygnus Loop.
%We find no exceptional abundance variations % in
%at the high spatial resolution of these
%data.
%measure normal interstellar abundances in
\end{abstract}

\section{Introduction}
Supernova remnants fundamentally
drive the cycle of matter and energy in the interstellar medium.
Supernova remnants (SNRs)
mix newly-formed elements into the interstellar medium (ISM),
and they may determine whether those elements remain gaseous or
enter the solid phase in the form of dust grains.
SNRs drive mass exchange between the different phases of the ISM.
Each SNR individually consists of an expanding hot interior, which
can evaporate surrounding cold clouds.  The compression
of the ambient medium behind the SNR blast waves sets the
stage for future generations of star formation.
The interaction between the ISM and the stars in it is reciprocal.
The stellar progenitor is responsible for
processing the local medium, which
alters subsequent evolution of the shock front,
while the immediate environment profoundly influences
the evolution of an individual SNR.

\begin{figure}
\includegraphics[width=6in]{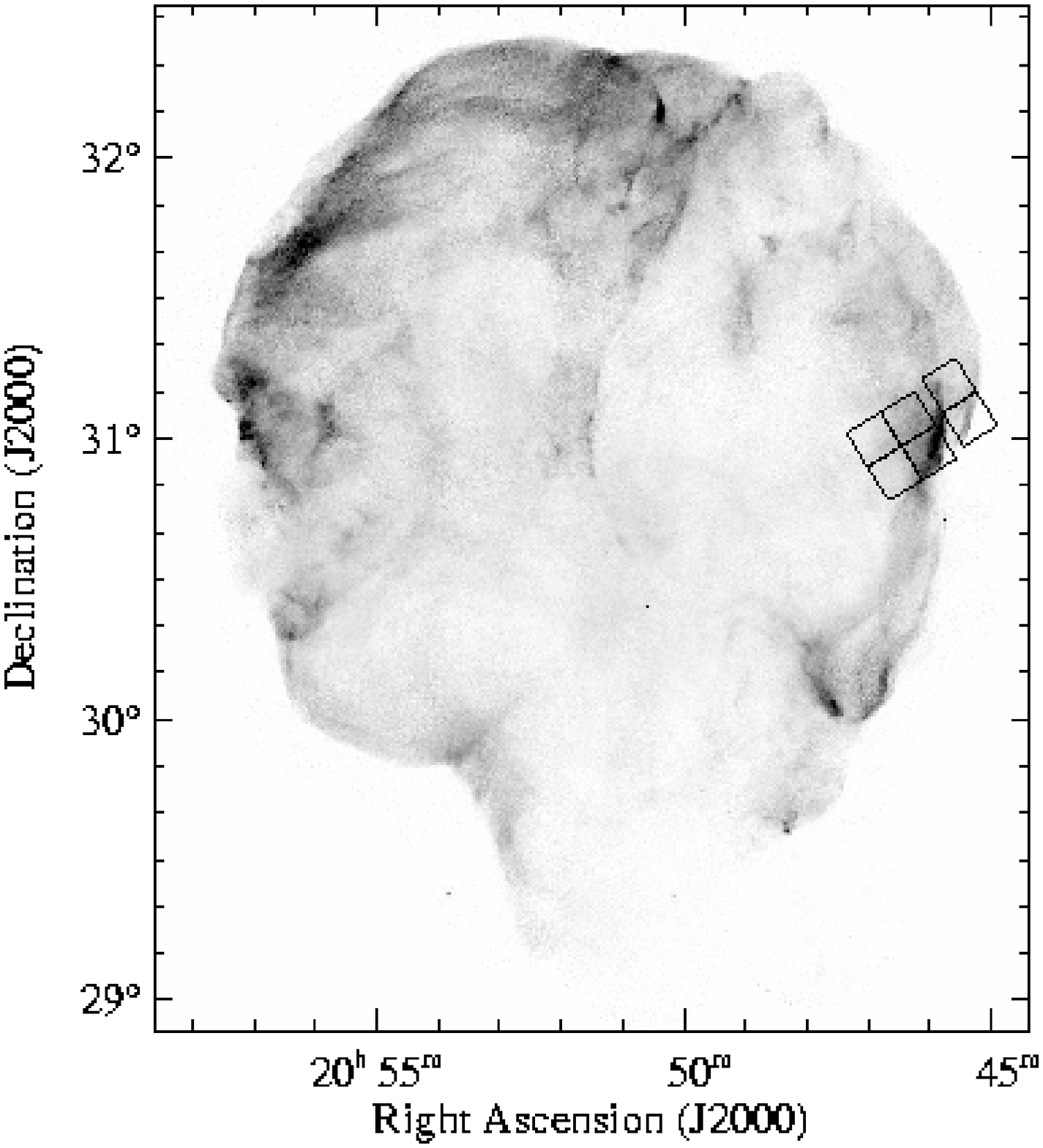} 
\caption{\label{fig:hriacis}
The context of these 31-ks \chandra{} observations 
is illustrated with the ACIS field of view overlaid on the
{\it ROSAT} HRI mosaic of the Cygnus Loop \citep[completed with
later observations]{Lev97}.
Interactions of
the blast wave with large interstellar clouds produce the brightest
X-ray regions.  Forward blast wave propagation in the very dense cloud medium
and the development of reflected shocks, which further heat and compress
previously-shocked gas, together enhance X-ray emission at these
locations.
}
\end{figure}

The  bright Cygnus Loop supernova remnant serves as an
ideal probe of its surroundings (Fig. \ref{fig:hriacis}).  
It is middle-aged ($\tau = 8000$ yr), 
with the extant ISM determining its subsequent evolution.
Because the Cygnus Loop is nearby (440 pc; \citealt{Bla99}), 
high spatial resolution, 
high signal-to-noise observations with the {\it Chandra X-ray Observatory}
(\chandra) reveal hydrodynamical evolution of the blast wave
and directly probe the inhomogeneous ambient ISM.
Large clouds and
a smooth atomic shell together define the boundary of the
cavity the SNR's progenitor created \citep{Lev98},
and shocks associated with both the cloud and shell
produce X-rays, which we analyze here. 

\begin{figure}
\includegraphics[width=6in]{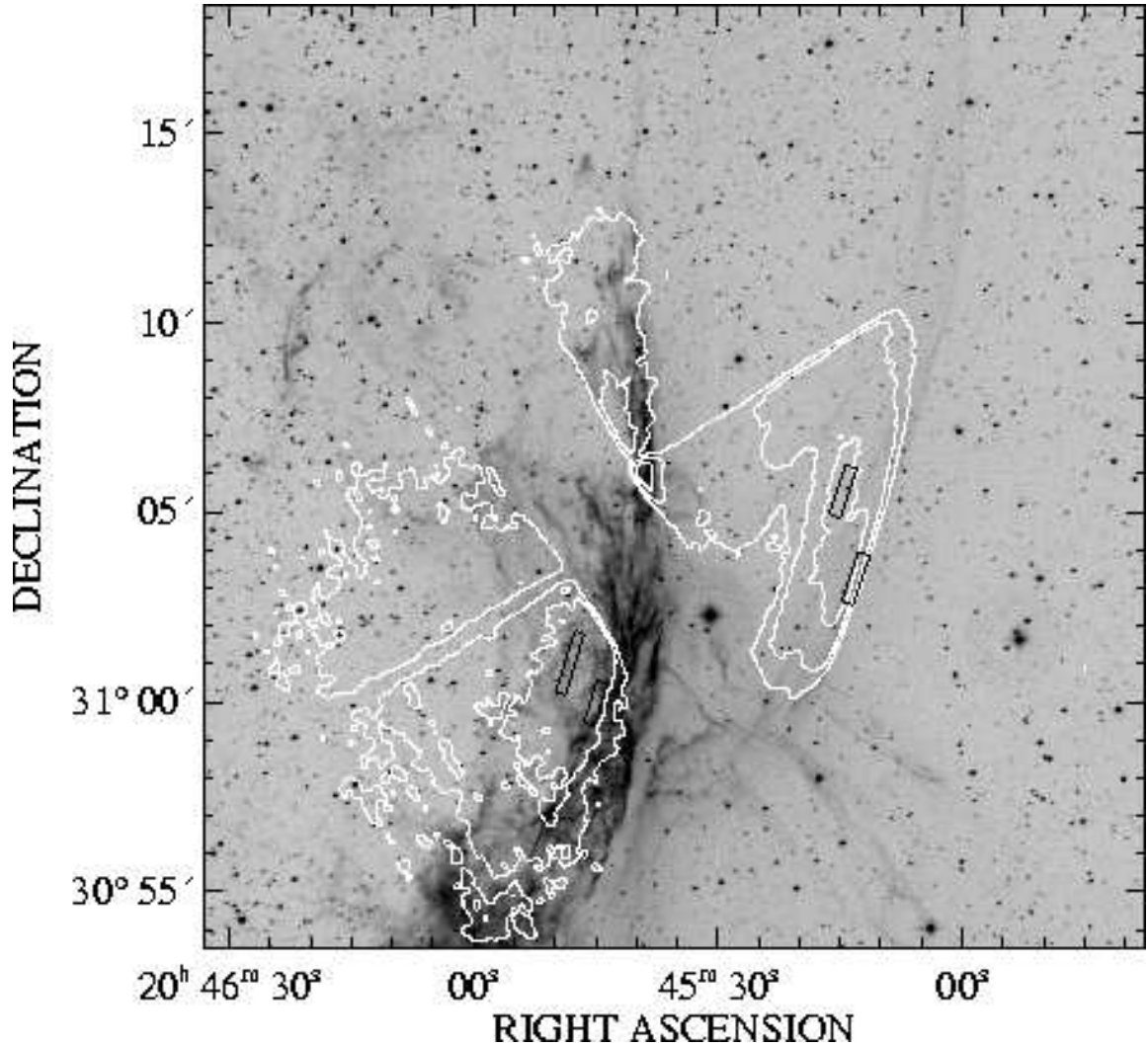} 
\caption{\label{fig:opt}
H$\alpha$ image of the western limb of the Cygnus Loop,
on a logarithmic scale \citep{Lev98}.  
Fully radiative shocks produce
the bright emission at the center of the field,
while Balmer-dominated shocks
trace the blast wave at the faint filaments toward the west.
Contours of \chandra{}
X-ray emission (white), also scaled logarithmically,
and the spectral extraction regions (black rectangles) are
overlaid.  Some of the CCD boundaries are evident in the X-ray
contours.
}
\end{figure}

The brightest regions of the Cygnus Loop in optical and X-ray emission,
such as the western and northeast limbs, 
are sites of interactions between the SNR blast wave and large
interstellar clouds, which extend over scales of 10 pc.
The blast wave is decelerated in the dense clouds.
The shocks that advance in the clouds  
are radiative, and their cooling gas
emits strongly in many optical lines, including hydrogen Balmer
lines,  [\ion{O}{3}], and [\ion{S}{2}], with 
typical temperature $T\sim 10^4$ K.
Figure \ref{fig:opt} illustrates the Cygnus Loop's characteristic
bright H$\alpha$ emission due to these slow cloud shocks at the western
limb.
Shocks that are reflected off the cloud
surfaces propagate back through previously-shocked material, further
heating and compressing it.  These
reflected shock regions exhibit enhanced X-ray emission \citep{Hes86},
with characteristic temperature $T\sim 10^6$ K. 

The extreme western limb also includes a portion of
the primary blast wave.  \citet{Ray80} first identified
its  characteristic faint filaments in  H$\alpha$ images and showed
that H$\alpha$ and H$\beta$ are 
the only strong lines in a filament's optical spectrum.
These ``non-radiative'' shocks
excite Balmer line
emission through electron collisions in pre-shock gas that
is predominantly neutral \citep{Che78,Che80},  
marking regions where the
gas is being shocked for the first time.  
The excitation is confined to a
narrow zone immediately behind the shock front, 
so these filaments
delineate the outer edge of the blast wave.
This true boundary of the SNR is 
now located in 
the neutral shell of material 
at the edge of the cavity the progenitor star created.

The western limb is the archetype of cloud--blast-wave encounters
in the Cygnus Loop.  
The large obstacle is detected directly in molecular observations \citep{Sco77}.
More importantly,
at this location
we view the collision with
the cloud nearly edge-on \citep{Lev96}.  The
stratification of the forward and reflected 
shock regions is therefore clear, and 
the simplification to a one-dimensional hydrodynamical
problem on small resolved angular scales is reasonable.
The X-ray data we present here also reveal
the similar
interaction of the primary blast wave with the cavity shell,
allowing comparison of these effects in different interstellar conditions.

\section{Observations and Data Reduction\label{sec:obsv}}

We obtained a 31 ks exposure of the western limb of the Cygnus Loop
on 2000 March 13 and 14 with the \chandra{} 
Advanced CCD Imaging Spectrometer 
(ACIS).  
(See \citealt{Wei00} and the Proposers' Observatory Guide at
http://asc.harvard.edu/udocs/docs/docs.html
for more information on \chandra{} and ACIS.) 
The total field of view is approximately $25\arcmin \times 17 \arcmin$,
with spatial resolution around $1\arcsec$ and 
simultaneous spectral resolution
$E/\Delta E \approx 12 $. 
We illustrate the  X-ray context of these
observations in Figure \ref{fig:hriacis},
marking the fields of view from the six \chandra{} detectors
on the \rosat{} High Resolution
Imager (HRI) soft X-ray mosaic of the entire SNR.  
The presence of a large
interstellar cloud enhances emission and 
indents the spherical blast wave here.

\begin{figure}
\includegraphics[width=6in]{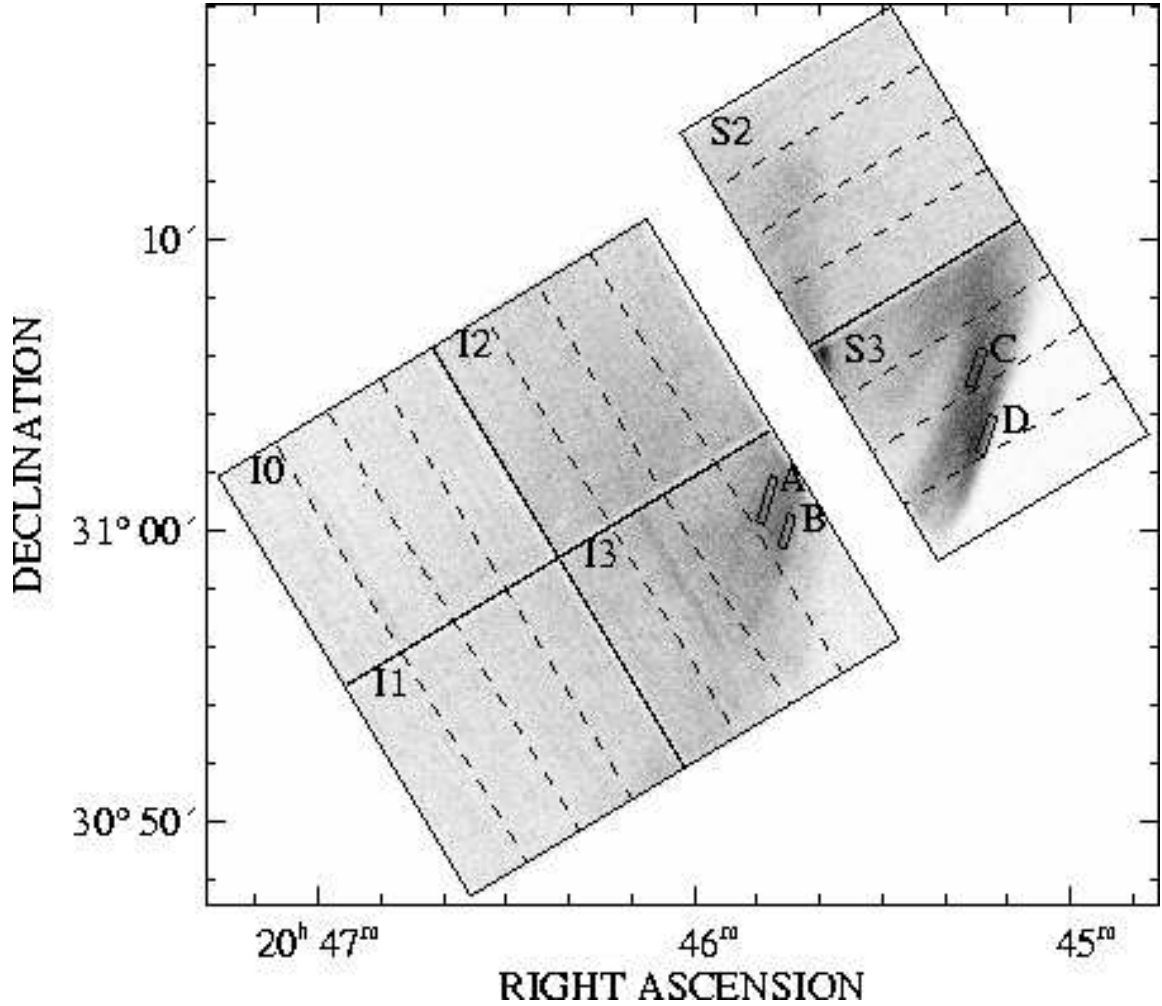} 
\caption{\label{fig:imgnodes}
\chandra{} ACIS image of the western limb.
This 0.1--8 keV image has been smoothed by FWHM$=3\arcsec$
and is scaled
linearly from 0 (white) to 7 (black) total 
counts per $1\arcsec \times 1\arcsec$ pixel.  
The individual detectors are identified by name
and marked with solid outlines. The
approximate node boundaries are indicated (dashed lines).  
Rectangles (A--D) identify the regions for
spectral analysis.
}
\end{figure}

Figure \ref{fig:imgnodes} contains the total ACIS 0.1--8 keV image binned
by a factor of two into $1\arcsec \times 1\arcsec$ pixels,
with the detectors identified.
This image is not corrected for existing large
variations in detector sensitivity.
The S3 CCD is oriented with its charge-transferring electrodes facing
away from the incident X-rays, which greatly increases
the soft X-ray sensitivity of this ``back-illuminated''
detector.  The remaining five CCDs are ``front-illuminated,''
having readout electronics that face the incident photons
and therefore absorb some of the soft X-rays from the astronomical
source.
The increased number of counts in the 
back-illuminated S3 CCD is the result of these 
differences, not varying emission intrinsic to the
source.
Within a given detector, 
each of the four readout amplifier nodes
has a distinct spectral response, which produces
smaller-magnitude variations; 
these node boundaries are also marked.

We reprocessed all data from original Level 1 event files,
removing the 0.5-pixel spatial randomization that is included in standard
processing.  
(See the \chandra{} Science Center at 
http://cxc.harvard.edu/ for details about \chandra{}
data and standard processing procedures.)
The latest
gain files (from the \chandra{} calibration database, version 2.7,
at http://cxc.harvard.edu/caldb/)
were used to calibrate the back-illuminated S3 
detector.  
During the first few months of \chandra{} operations, radiation
damaged the front-illuminated devices, increasing their
charge transfer inefficiency, consequently diminishing their
spectral resolution and sensitivity.
To mitigate these effects, 
we used the software and technique of \citet{Tow00},
applying their 
response matrices for energy calibration.
We included only good events that 
do not lie on node boundaries, where discrimination of cosmic rays
is difficult.
We examined the lightcurves of background regions and found no
significant flares, so we did not reject any additional data
from the standard good-time intervals.

We combined three energy-selected images to create the
false-color composite (Fig. \ref{fig:clr}).
Total counts 
in the 0.3--0.6, 0.6--0.9,
and 0.9--2.0 keV energy bands are displayed in red, green, and blue,
respectively, and the individual images
have been binned by a factor of two and smoothed by FWHM$=14\arcsec$.
The individual energy images are not corrected for sensitivity
variations within each detector or across the field.
Thus, some of the color variation is not due to
intrinsic spectral variation.  Notably,
because of its distinct spectral sensitivity as 
a function of energy, the correspondence
between color and spectral shape in the back-illuminated S3 detector is
different from that of the other five front-illuminated CCDs, 
so the S3 detector is displayed separately.
Most importantly, the S3 detector is more sensitive at very soft
X-ray energies.
The component red, green, and blue images 
are scaled linearly from 0 to 2, 4, and 1.5, respectively, 
in the  front-illuminated detectors,
and  to 5, 3, and 1, respectively,
in the back-illuminated detector,
in units of 
$10^{-7} {\rm photons \, cm^{-2} \, arcsec^{-2}}$.

\section{Spectral Modeling}
The false-color image (Fig. \ref{fig:clr}) illustrates significant spectral variation
on small spatial scales within the Cygnus Loop.
We identified several outstanding features
(indicated in Fig. \ref{fig:imgnodes})
that correspond to specific physical circumstances in the
context of interactions between the SNR shock and the inhomogeneous
ISM, as we demonstrate below.
We extracted spectra from these small (roughly $1\arcmin \times 15\arcsec$)
regions 
in order to isolate their particular spectral characteristics
and to avoid calibration variations that are significant
over larger areas.  
Within the bright emission of the I3 front-illuminated detector,
the spectrum of region A is relatively hard, and that of 
region B is extremely soft.
Immediately behind the primary blast wave observed on the S3 CCD,
region C is spectrally harder than region D.
Each spectral extraction was confined to a single read-out node
of a detector, to avoid significant 
calibration variations that occur across node boundaries, 
 and in each case, background emission was measured 
in an off-source region of the same CCD.
Data were grouped into bins with a minimum of 20 counts,
so $\chi^2$ statistics are appropriate in the model fitting.  

We fit the spectra satisfactorily with thermal equilibrium plasma models.
Specifically, we used the {MEKAL} 
model \citep{Mew85,Arn85,Mew86,Kaa92},
with updated Fe L calculations \citep{Lie92}
in {XSPEC} \citep{Arn96}.  
We used data in the energy range 0.3--2.0 keV from the back-illuminated
detector and restricted the energy range to 0.4--2.0 keV from the
front-illuminated CCD.  The detector calibration uncertainties determined
the lower energy bound, and the lack of significant emission above
2 keV in all instances set the upper bound.
In all cases, we allowed temperature and
absorbing column density to be free parameters.
We constrained the latter $N_H \ge 4.7\times 10^{20} \psc$,
the column density that corresponds to the average visual
extinction to the Cygnus Loop \citep{Par67}, 
which we assume is in the foreground,
allowing for additional intrinsic absorption or variation as a function
of position.
Regions C and D do not require any additional absorption, so
we fixed $N_H = 4.7\times 10^{20} \psc$ in their final spectral fits. 
Figure \ref{fig:spec} shows the data, best-fitting models, and residuals
for the four regions, and Table \ref{tab:specfit} lists the model parameters,
corresponding emission measure, and count rates.

With one exception, a single temperature component is sufficient
to describe the data with 99\% confidence.
In fitting the spectrum of region B, however, the inclusion of a second
thermal component is statistically significant.  As we describe
below, this extraction does not cover the cooler cloud shock
exclusively, but also contains an additional contribution
along the line of sight from the hotter reflected shock, to which
the detector is more sensitive.  In all other cases, however,
the addition of any second spectral component does not significantly 
improve the model fits.

In general, sub-solar oxygen abundance is
necessary to fit the data, where $({\rm O/H})_\odot = 8.51\times10^{-4}$
\citep{And89}.  
Because the characteristic temperatures we measure 
in the Cygnus Loop are very low 
compared with younger SNRs 
($kT \le 0.2$ keV), 
these spectra are particularly
sensitive to oxygen, which accounts for nearly all the emission
from 0.5--0.7 keV and represents a significant fraction of the detected
photons.  
We examined the harder and softer portions of the spectra separately
to confirm that they differ only in
normalization and not temperature or ionization state, for example, 
so adjusting the oxygen abundance is reasonable.
In the final modeling, we fixed the oxygen abundance
at the best-fitting value for each detector, 
determined in
the fitting of at least four spatially distinct extractions where the
abundance was allowed to be a free parameter.
We use ${\rm O/H} = 0.53({\rm O/H})_\odot$ in the I3 CCD and
${\rm O/H} = 0.44({\rm O/H})_\odot$ in the S3 detector. 
These values are similar
to recent photospheric abundance measurements \citep[e.g., ][]{Hol01}.

We do not measure any systematic trend of abundance as a function of distance
from the shock front, which indicates that we are not sensitive
to the immediate effects of depletion and grain destruction in
interstellar shocks.
The offset, however, between the front-illuminated and back-illuminated
CCDs is due to a difference in the relative soft-energy 
calibration of these detectors, not a physical variation over
the spatial scale we investigate.  
We measure this calibration offset in a cloud interaction
region that covers  the S3 and I3 CCDs contiguously.
The common physical conditions of this spatially-extended emission
produce the same detector-specific abundance difference.
Unfortunately, these observations are not strongly dependent on
the abundances of elements that grain depletion affects most greatly.
These data are not at all sensitive to silicon abundance,
for example.  At some of the higher observed temperatures ($kT \approx 0.2$ keV),
we could measure depletion of iron at the 10\% level, but because these
hotter regions represent older shocked material, they are not physically
revealing.

In addition to the broad oxygen complex, we note several other features 
in the spectra.  \ion{Mg}{11} produces the emission 
near 1.35 keV in spectra from regions A and B.  In the spectra of
regions C and D, \ion{Ne}{9} transitions near 0.92 keV are obvious.
In our final spectral analysis, we use the 
MEKAL model because it reproduces the prominent 
Ne emission significantly better than
other equilibrium  models.
Qualitatively, the absence of \ion{Fe}{17} emission 
around 0.73 and 0.83 keV emphasizes the relatively low temperatures of
all the observed X-ray--emitting regions; these lines produce significant
emission when $kT \ge 0.5$ keV.

We considered non-equilibrium ionization in the spectral modeling,
which may be expected to be physically relevant to the conditions
in the Cygnus Loop.  
Collisional ionization is not immediate following the
shock, so the elements are initially
under-ionized with respect to their equilibrium values.
The ionization parameter, $n_e t$, 
characterizes the scale of equilibration,
where $n_e$ is the
initial electron density  and $t$ is the
time elapsed since the passage of the shock.
Ionization equilibrium usually occurs when 
$n_e t \ge  3\times 10^{11} {\rm \,cm^{-3}\,s}$, depending on the 
element and its equilibrium state.
The best fitting non-equilibrium ionization models tend toward equilibrium
solutions, with 
$n_e t \gg 10^{12}  {\rm \,cm^{-3}\,s}$.
Furthermore, these non-equilibrium models
do not statistically improve the spectral fits,
so we do not consider them further here.  

On physical grounds, we would expect non-equilibrium conditions to be
marginally relevant.  Using the densities and timescales calculated 
below (\S \ref{sec:physcond}), 
we find $n_e t \sim 10^{12} {\rm \,cm^{-3}\, s}$,
which is somewhat higher than previous results \citep[e.g.,][]{Ved86,Miy94}.
At the very soft X-ray energies of the Cygnus Loop, however, no instrument
has the spectral resolution and sensitivity on small spatial scales 
required to measure the non-equilibrium conditions directly and accurately. 
The spectra of a fully-equilibrated lower-temperature medium and
the non-equilibrium state of a higher-temperature plasma remain  
indistinguishable.
Thus, the adoption of the better-fitting equilibrium models in
this analysis is justified.

In this inhomogeneous region, high spatial resolution is
essential to derive physically meaningful conclusions from
the spectral analysis.
For comparison, we extracted the spectrum from a $2\arcmin$
circular region of the bright emission observed with the I3 detector.
None of the models we applied, including multiple temperatures
and non-equilibrium, fit this spectrum adequately because
it covers such a range of physical conditions.
We note that similar to analysis of {\it ASCA} spectra of
other locations in the Cygnus Loop at
comparable spatial resolution, 
extreme abundance variations are statistically preferred \citep{Miy99}.
The sensitive \chandra{} data, however, make clear that 
such exceptional models are still unacceptable, 
and they emphasize that spatially-integrated spectra
do not directly measure genuine abundance variations.

\section{Physical Interpretation\label{sec:physcond}}

\subsection{Cloud Shock}
These \chandra{} data reveal the propagation of the
SNR blast wave through the inhomogeneous ISM at the western
limb of the Cygnus Loop.  Over most of this region,
we observe the interaction with a large
interstellar cloud at the boundary of the pre-supernova cavity
\citep{Sco77}.   
When the blast wave encounters the cloud, three characteristic regions
become important: the decelerated forward shock in the cloud, 
a reflected shock,
which propagates back through the previously-shocked ambient ISM, 
and the singly-shocked SNR interior \citep{Hes86}.
Because we view the plane of the shock nearly edge-on, 
the stratification of the spatially distinct emission regions is preserved.
(See fig. 7 of \citealt{Lev96} for a cartoon
illustration of the geometry of this region.)

\begin{figure}
\includegraphics[width=3.5in]{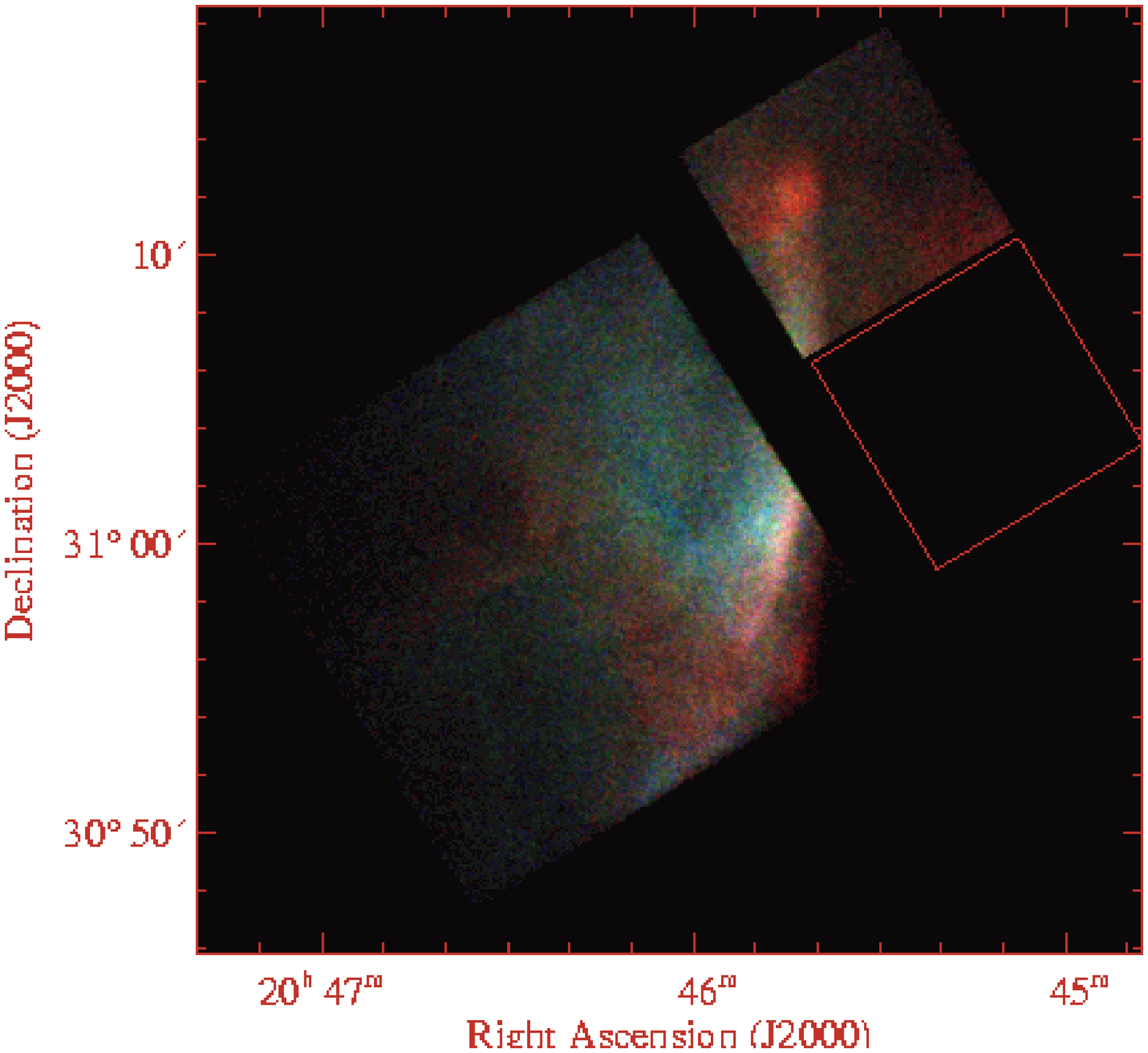} 
\includegraphics[width=3.5in]{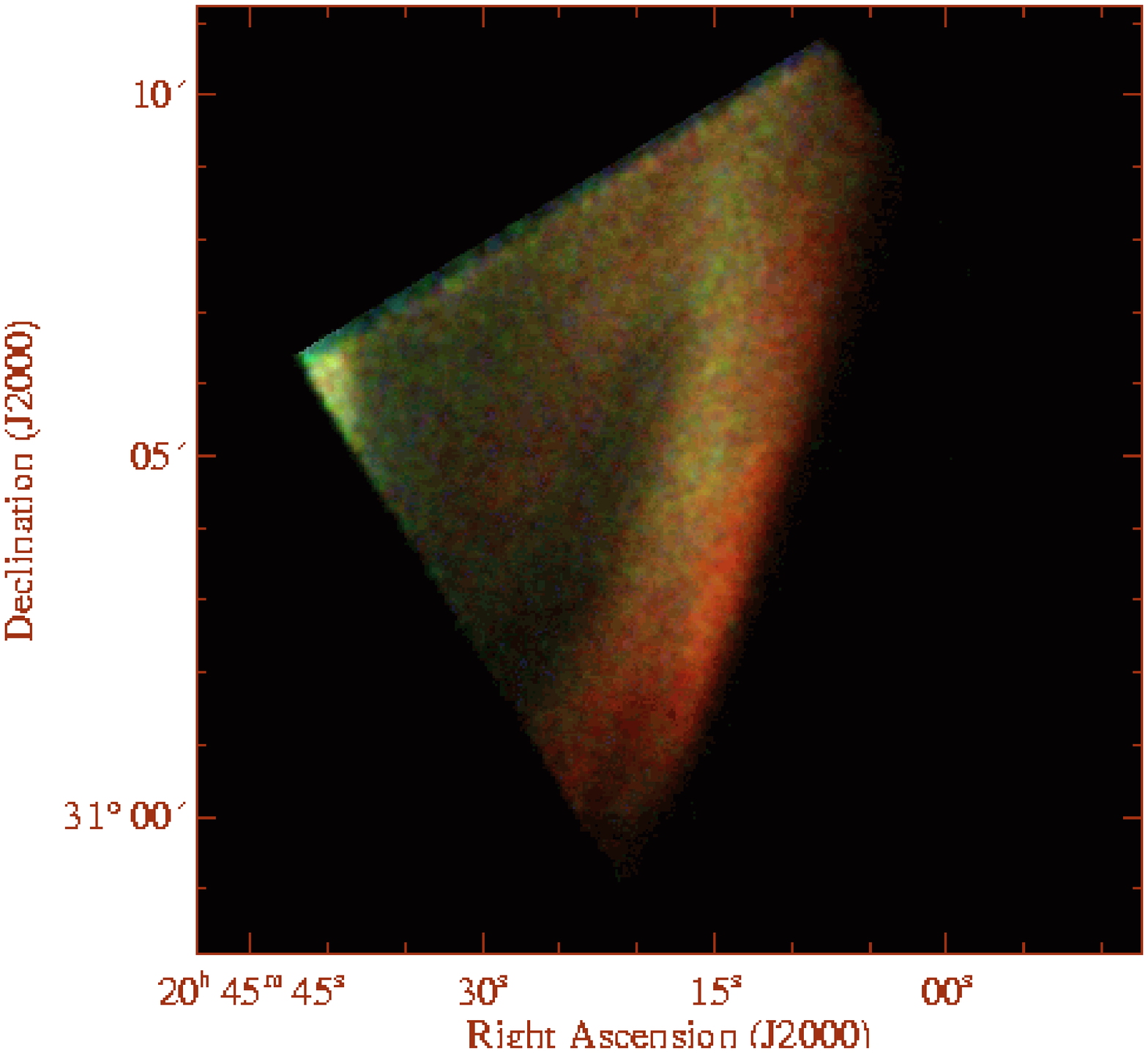} 
\caption{\label{fig:clr}
These false-color images reveal varying physical
conditions over the field of view.  Total counts 
in the 0.3--0.6, 0.6--0.9,
and 0.9--2.0 keV energy bands are displayed in red, green, and blue,
respectively, and the individual images
have been binned by a factor of two and smoothed by FWHM$=14^{\prime\prime}$.  
The red square on the left indicates the location of the S3 detector
 (right),
which is scaled differently because of its greater
sensitivity at lower energies. (See \S \ref{sec:obsv} for details.)
The softest (reddest) emission arises behind 
the decelerated blast wave as it propagates through 
the dense cloud and shell material.  Reflected shocks produce
corresponding hotter emission behind these slow shocks.
}
\end{figure}

\begin{figure}
\includegraphics[angle=-90,width=3.0in]{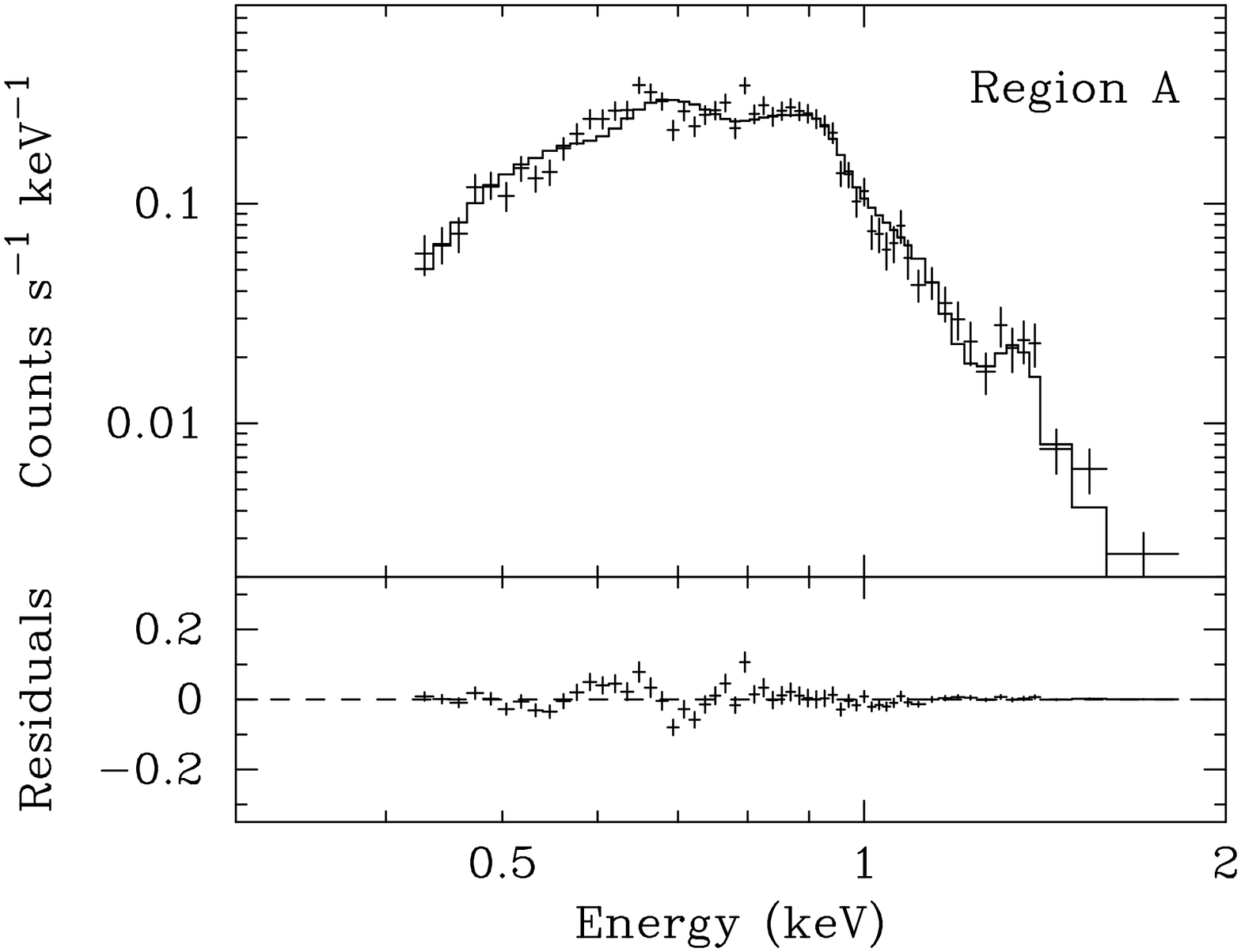} 
\includegraphics[angle=-90,width=3.0in]{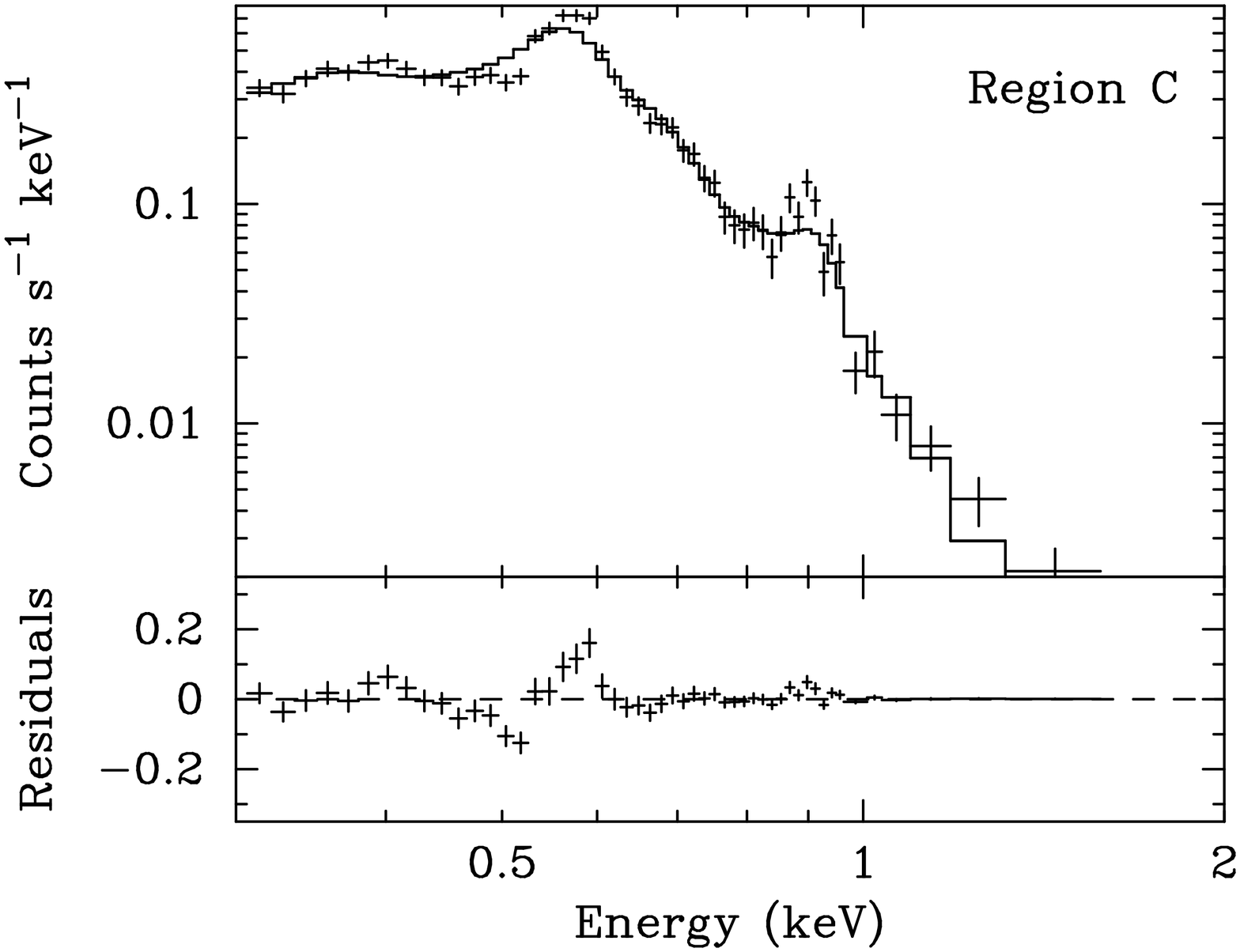} 

\includegraphics[angle=-90,width=3.0in]{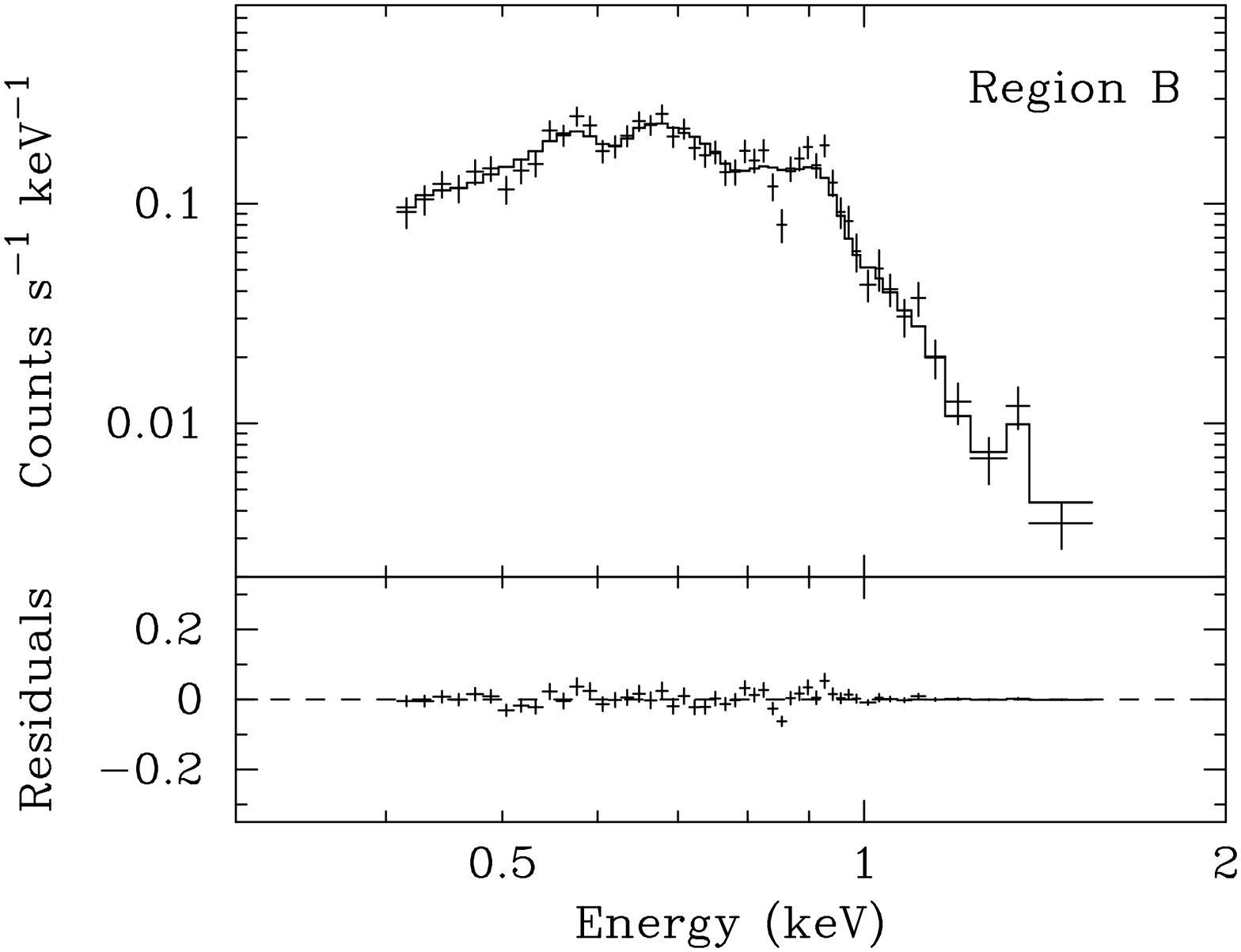} 
\includegraphics[angle=-90,width=3.0in]{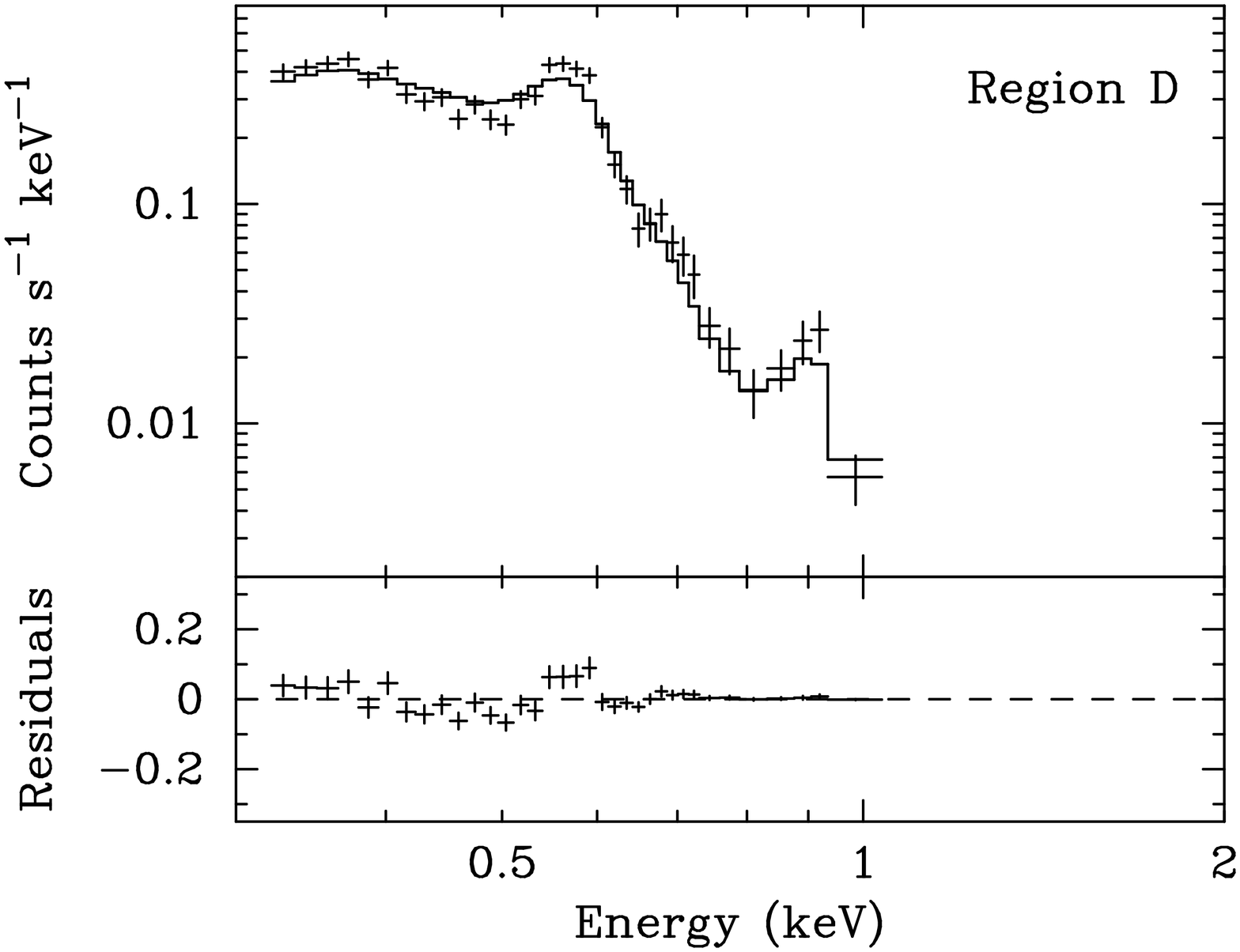} 

\caption{\label{fig:spec}
Spectral data and model fits from the regions
noted in Figure \ref{fig:imgnodes}.  Table \ref{tab:specfit} contains the 
best-fitting model parameters.  Equilibrium plasma models
characterize these spectra, and we measure
significant temperature variations among the different regions.
}
\end{figure}

The distinct, narrow, pink feature 
in Figure \ref{fig:clr}, which extends over several arcminutes
from north to south, directly illustrates the shock front in
the cloud.  
Even though the cloud shock is slower than the
undisturbed blast wave and will therefore
not produce such {\em hot} plasma,
the shock in the denser cloud produces very {\em bright}
X-ray emission; for a given volume, emitted flux is 
proportional to $n^2$, where $n$ is density.
The emission from the reflected shock region is also enhanced
with respect to the rarefied singly-shocked SNR interior.  
The intense X-ray emission of this interaction
was previously evident at the spatial resolution
of the \rosat{} HRI observations, but without simultaneous
spectral information, the decelerated cloud shock and
the harder reflected shock could not be distinguished.
In these \chandra{} data, however, 
we uniquely identify both constituents 
of the X-ray enhancement.
The location of the pink emission further identifies it 
as the cloud shock.  It appears ahead of the harder
reflected shock, and it is aligned with the onset
of a radiative cooling zone observed optically (Fig. \ref{fig:opt}).

The peak emissivity of \ion{O}{6} occurs
around the temperature of the shocked cloud, 
$T = 3.5 \times 10^5$ K,
and oxygen is the dominant coolant at this temperature. 
Far-ultraviolet maps of the Cygnus Loop 
that are sensitive to the \ion{O}{6}
doublet at $\lambda\lambda$1032, 1038 show strong emission at the
western limb and are generally correlated
with the X-rays \citep{Bla91,Ras92}, consistent with
\ion{O}{6} production in cloud shocks.

Because of the nearly edge-on geometry of this
interaction, projection effects are not significant in this case,
but they are present and explicable.
For example, the emission slightly ahead (west) of the cloud
shock yet still within the I3 CCD is due to another portion of
the blast wave, which is likely on the near side of the cloud.  The
extremely soft (red in Fig. \ref{fig:clr}) patch near
$\alpha = 20^{\rm h}45^{\rm m}40^{\rm s}$, $\delta = 30^\circ52\arcmin$ (J2000)
of this region is associated with a fully radiative shock, detected
at optical wavelengths in H$\alpha$ and [\ion{O}{3}]$\lambda5007$
lines.  The relationship between multiple projections
of the  extended shock plane is expected when it encounters
the non-uniform surface of an interstellar cloud.
Another consequence of projection effects is that the spectral extraction
of region B includes some contribution from the reflected shock in
addition to the cloud shock along the line of sight.
Qualitatively, this effect produces the pink cloud shock:
the softest (red) emission is most significant, 
and combined with secondary harder (green and blue) contributions of the
reflected shock along the line of sight, 
the net result is pink in this image. 
\chandra's increasing effective area with energy up to 1 keV emphasizes
this effect in the spectral extractions, and 
the spectral model of this region requires both the very soft
emission of the slow forward shock and the hotter emission
of the reflected shock that we measure in the spectrum of region A.

Distinguishing these two spectral components, we quantify
the relationship between the shock velocities and density contrast
of the cloud and cavity regions.
For standard interstellar abundances, the post-shock temperature and
cloud shock velocity $v_{c}$,
are related by 
$T = 1.4\times 10^5 (v_c/100 {\rm \,km \,s^{-1}})^2$ K.  The measured
temperature $T = 3.5$ (+6.7, -1.9) $\times 10^5$ K corresponds to
$v_c = 150$ (+120, -40) ${\rm\, km\,s^{-1}}$.
We solve the one-dimensional conservation equations of 
mass, energy, and momentum 
over both the cloud and reflected shock fronts, 
assuming constant pressure across the contact discontinuity that
develops between the singly-shocked ambient medium and the reflected
shock region.  (See \citealt*{Hes94}, for example.\footnote{Note the
typographical error in the equation on page 739 of \citet{Hes94}, 
which should read
$${{1}\over{y}} = 1 - {{3}\over{4\eta}}(1-\beta).$$})
Because the region of interest is small compared with the
total size of the SNR, we reasonably assume that the initial
ambient pressure is constant. 
While we emphasize that the limb-brightened morphology of the Cygnus Loop
is not a consequence of SNR evolution in a homogeneous medium,
the Sedov similarity solution demonstrates that pressure behind the
blast wave as a function of radius, $R$, over 
the observed scale $\Delta R$, is
approximately constant here, with  $\Delta R/R \approx 10^{-2}$.
In this case, the observed temperature difference between 
the cloud and reflected shock regions
corresponds to a density contrast of 11 between the original cloud and
the ambient cavity medium, and the original blast
wave in the rarefied cavity had $v_{bw} = 310 {\rm \, km \, s^{-1}}$.

We use the reflected shock spectrum to determine the densities
because its normalization, the relevant model parameter, 
is better constrained than the normalization of the cloud shock.
We assume the emitting region extends 5 pc along the line
of sight, the depth of the SNR at this radius.
The reflected shock model then indicates an initial cavity
density $n_0 = 0.4 {\rm \, cm^{-3}}$, and cloud density
$n_c = 5 {\rm \, cm^{-3}}$. 
The thermal pressure in the fully-ionized reflected region
$P=2.0\times 10^{-9} {\rm \, dyne \, cm^{-2}}$.
According to the solution of the reflected shock equations,
this should be a factor of 2.65 greater than the pressure
behind the ambient blast wave, so these data imply 
$P_{bw} = 7.5\times10^{-10}{\rm \, dyne \, cm^{-2}}$.
We compare these derived physical quantities with the results from
the shell region and previous work below (\S \ref{subsec:shell}).

\subsection{Shell Shock\label{subsec:shell}}
At the far west, observed in the S3 detector, the
SNR blast wave  encounters the atomic shell that is
a consequence of progenitor's processing of the ISM \citep{Lev98}.
The softest spectrum (region D) is near the cloud, where the density
(and therefore shock deceleration) are greatest. 
The temperature $kT = 0.12$ (+0.004, -0.003) keV corresponds to 
$v_s = 310$ (+7, -4) ${\rm \,km\,s^{-1}}$
in the shell.
Behind the blast wave and farther from the shell, the 
reflected shock component enhances and hardens the emission 
to $kT = 0.16$ keV.
Given the observed curvature of the shock front, 
we do not expect the shell to be uniform over the entire field of view, 
but we roughly estimate the density contrast between the shell and the 
ambient medium assuming that these extracted spectra are
the reflected and forward components of the same interaction.
The observed temperature ratio of 1.3 requires  a density
ratio of 1.5 between the original 
shell and the rarefied interior, and $v_{bw} = 340 {\rm \, km \, s^{-1}}$
for the original blast wave.

Assuming the thin ($54\arcsec$) shell observed in X-rays
is the spherical shell of the SNR of radius 11 pc,
the emitting volume of the 
extracted spectrum extends $6.9\times 10^{18}$ cm along the line of sight.  
Thus, the original shell density $n_s = 0.6 {\rm \, cm^{-3}}$, and
in the cavity, $n_0 = 0.4 {\rm \, cm^{-3}}$ before the passage of 
the strong blast wave, in  agreement with the
value obtained from modeling the cloud shock region.
The density and shock velocity ratios in the shell region indicate
a somewhat higher blast wave pressure:
$P_{bw} = 1.1\times 10^{-9} {\rm \, dyne \, cm^{-2}}$.

The errors are large in calculating the cavity and blast wave parameters
from both the cloud shock and shell shock regions.
In the former case, we more certainly identify the
cloud and reflected shock regions.  Disadvantageously, however, 
the computed cloud density
is sensitive to the intensity normalization of the plasma model,
but at such soft X-ray energies, we cannot measure the normalization
(or density) with \chandra{} to better than a factor of 10 in this
region that includes multiple temperature components.
In the simpler case of the shell shock, the errors in the density
measurement are only a factor of $\sqrt{2}$.  The question here,
however, is whether the slightly hotter region C is a clear example of 
a reflected shock.  Given these uncertainties, 
the similarity of these results from the two regions
suggests that they both roughly characterize
the pre-existing and present ISM conditions.  

Both the cavity and cloud densities we find are comparable to 
previous X-ray and optical measurements, after
correcting for the revised distance, where appropriate
\citep[e.g., ][]{Cox72,Fes82,Ku84,Ray88,Hes94}.
These results are also similar to 
distance-corrected global measurements of $P_{bw}$.
Ku et al.'s (1984) \nocite{Ku84} 
X-ray observations of the
entire Cygnus Loop yield
 $P_{bw} \approx 7\times 10^{-10} {\rm \, dyne \, cm^{-2}}$,
for example.
We note that considering the additional heating the
reflected shock provides, 
we derive a somewhat lower
value of $v_{bw}$ than others find from X-ray data.
The reason for this discrepancy is that previous work has
attributed $kT \approx 0.2$, characteristic of the
brighter regions,
to the forward blast wave alone, 
for $v_{bw}\approx 400 {\rm \, km \, s^{-1}}$.

\subsection{Instabilities}

At the present spatial and spectral resolution, 
non-equilibrium conditions 
do not affect the measurable X-ray properties significantly.
Morphologically, the cloud shock front is evident 
as a distinct feature in the energy map, having 
width less than $10\arcsec$,
while extending over a length of several arcminutes.
Quantitatively, the spectra of the various regions, including
those at the extreme western limb as well as the bright central
region, correspond to realistic physical conditions in
the ISM in the context of the reflected shock model.  

The cloud shock temperature is uniform on resolved scales,
which implies that shear flow or other instabilities are not important
at this shock front, thereby justifying this simplified one-dimensional
treatment.  While more complex hydrodynamic models have
been proposed to interpret cloud-shock interactions, they are
not required in this case.  Turbulent mixing layers, for
example, are predicted to cascade down from
large to small scales \citep{Beg90,Sla93}.
If these mixing layers are present, their 
length scale must be much smaller than
the arcsecond ($10^{16}$ cm) scale we resolve here.  We do not
detect the strong ionization and temperature disequilibrium these
models predict, despite 
the early stage of the encounter that could establish 
the shear flow such models require that is exhibited here. 

These conclusions based on the X-ray data broadly
agree with other observations of the Cygnus Loop.
At optical wavelengths, however, high sensitivity and
high spatial resolution reveal a great deal of additional structure,
much of which may be a consequence of 
variations in the pre-existing cloud.  The shell interaction
at the far-western limb, however, appears to be extremely uniform in
all observations.  It exhibits smooth Balmer filaments over its length,
which have not developed into radiative shocks,
for example.  The combined data therefore imply that 
the shell wall is uniform, at least on 
the arcsecond  scales that are resolved. 

\section{Conclusions}
These \chandra{} observations reveal the physical conditions
of the prominent western limb of the Cygnus Loop.
The interaction of the primary blast wave and a large interstellar
cloud enhances X-ray emission, where both the decelerated shock 
in the dense cloud and the development of a reflected shock 
are important.  
The primary blast wave currently propagates through a shell
of material that marks the boundary of the pre-supernova cavity.
Here, too, the X-rays reveal the inhomogeneity of the ISM
and clearly distinguish the Cygnus Loop from supernova expansion
in a uniform medium.
These X-ray data uniquely allow direct and simultaneous measurement
of the multiple shock components that are present, which
we quantify with spectroscopy of several distinct regions.
These data demonstrate the 
necessity of high spatial resolution spectroscopy, and
they directly reveal the hydrodynamic evolution 
of the supernova remnant on small physical scales.

The edge-on view of the shock propagation allows simplified
one-dimensional analysis.  
The blast wave has not yet passed beyond the cloud, so
this encounter is currently
at an early stage with respect to the large cloud.
It is, however, a well-developed interaction, in which 
we do not find significant evidence for instabilities or  
non-equilibrium conditions.
The western limb may be exceptional in this respect.
Morphologically, it appears to be much simpler than other
bright cloud--blast-wave interactions of the Cygnus Loop,
such as the eastern limb and the southeast knot, so
dynamical effects may still be important in these other regions.

\begin{acknowledgements}
We thank Paul Plucinsky and Glenn Allen of 
the \chandra{} X-ray Observatory Center
for their assistance.  We thank Chris McKee 
and the anonymous referee for their comments, which 
improved this work.  
N.A.L. appreciates 
many useful discussions with David Strickland about data processing.
Support for this work was provided by the National Aeronautics and
Space Administration through \chandra{} Award Number GO0-1121
issued by the \chandra{} X-ray Observatory Center, which is
operated by the Smithsonian Astrophysical Observatory for and on behalf of
the National Aeronautics and Space Administration under contract
NAS8-39073.
\end{acknowledgements}

\input tab1
\clearpage

\end{document}

%% file: tab1.tex
\begin{deluxetable}{lllllcc}
\tabletypesize{\scriptsize}
\tablewidth{0pt} 
\tablecaption{Spectral Model Fits\label{tab:specfit}}

\tablehead{
\colhead{Region}&
\colhead{Count Rate}&
\colhead{EM}&
\colhead{$N_H$}&
\colhead{kT}&
\colhead{O Abundance\tablenotemark{a}}&
%\colhead{normalization}
\colhead{$\chi^2/\nu$}\\
&\colhead{(cts s$^{-1}$)}&
\colhead{(cm$^{-6}$ pc)}&
\colhead{($10^{20} {\rm\,cm^{-2}}$)}&
\colhead{(keV)}&
\colhead{($Z_\odot$)}
}

\startdata
A& $0.140\pm0.002$ & $72^{+29}_{-22}$ & $24^{+4.0}_{-4.1}$ &
$0.18^{+0.008}_{-0.006}$ & 0.53 & 105/56 \\
B\tablenotemark{b}& $0.102 \pm 0.002$& $< 2\times10^5$ &
$11^{+2.1}_{-2.0}$ & $0.030^{+0.06}_{-0.02}$ & 0.53 & 66/46 \\
C& $0.194\pm0.003$ & $14\pm 0.7$ &
4.7\tablenotemark{c} & $0.16\pm0.003$ & 0.44 & 126/49 \\
D& $0.116\pm0.002$& $16 \pm 1.4$ &
4.7\tablenotemark{c} & $0.12^{+0.004}_{-0.003}$ & 0.44 & 74/33 \\ 
\enddata
\tablenotetext{a}{Oxygen abundance is fixed at $0.53Z_\odot$ and 
$0.44Z_\odot$ in the front-illuminated and back-illuminated detectors,
respectively, based on best-fitting variable abundance for
several spectral extractions in each case.}
\tablenotetext{b}{Also includes second component with fixed $kT = 0.18$ keV,
similar to region A.}
\tablenotetext{c}{Fixed parameter.}
\tablecomments{
%All spectra were fit with the {MEKAL} thermal plasma model in {XSPEC}.  
Errors indicate 90\% confidence limits
for two variable parameters.}
\end{deluxetable}

%tablenotemark{},tablenottext{}{}

%C& $0.194\pm0.002$ & $32^{+5.2}_{-2.9}$ &
%$5.4^{+1.2}_{-1.1}$ & $0.15^{+0.004}_{-0.003}$ & 0.31 & 127/48 \\
%D& $0.183\pm0.002$& $200^{+150}_{-130}$ &
%$12^{+2.6}_{-5.0}$ & $0.096^{+0.017}_{-0.007}$ & 0.31 & 80/37 \\ 

%% file: levenson.bbl
\begin{thebibliography}{}
\bibitem[Anders \& Grevesse(1989)]{And89}Anders E., \& Grevesse, N. 1989, Geochimica et Cosmochimica Acta 53, 197
\bibitem[Arnaud(1996)]{Arn96}Arnaud, K. A. 1996, in ASP Conf. Ser. 101,
Astronomical Data Analysis Software and Systems V, ed. G. Jacoby \& J. Barnes 
(San Francisco: ASP), 17
\bibitem[Arnaud \& Rothenflug(1985)]{Arn85}Arnaud, M., \& Rothenflug, M. 1985, \aaps, 60, 425 
\bibitem[Begelman \& Fabian(1990)]{Beg90} Begelman, M.~C., \& Fabian, A.~C.\ 1990, \mnras, 244, 26P 
\bibitem[Blair et al.(1991)]{Bla91}Blair, W.~P., Long, K.~S., Vancura, O., \& Holberg, J.~B.\ 1991, \apj, 374, 202
\bibitem[Blair \ea (1999)]{Bla99}Blair, W. P., Sankrit, R., Raymond, J. C., \& Long, K. S. 1999, \aj, 118, 942 
\bibitem[Chevalier \& Raymond(1978)]{Che78}Chevalier, R. A., \& Raymond, J. C. 1978, \apjl, 225, L27 
\bibitem[Chevalier, Raymond, \& Kirshner(1980)]{Che80}Chevalier, R. A., Raymond, J. C., \& Kirshner, R. P. 1980, \apj, 235, 186 
\bibitem[Cox(1972)]{Cox72} Cox, D.~P.\ 1972, \apj, 178, 143
\bibitem[Fesen, Blair, \& Kirshner(1982)]{Fes82} Fesen, R.~A., Blair, W.~P., \& Kirshner, R.~P.\ 1982, \apj, 262, 171
\bibitem[Hester \& Cox(1986)]{Hes86}Hester, J. J., \& Cox, D. P. 1986, \apj, 300, 675
\bibitem[Hester \ea(1994)Hester, Raymond \& Blair]{Hes94}Hester, J. J., Raymond, J. C., \& Blair, W. P. 1994, \apj, 420, 721 
\bibitem[Holweger(2001)]{Hol01}Holweger, H. 2001,
in ``Solar and Galactic Composition,'' ed. R. F. Wimmer-Schweingruber, 
(NY: AIP Press), in press (astro-ph/0107426)
\bibitem[Kaastra(1992)]{Kaa92}Kaastra, J. S. 1992, ``An X-Ray Spectral Code for Optically Thin Plasmas,'' Internal SRON-Leiden Report 
\bibitem[Ku et al.(1984)]{Ku84}Ku, W.\ H.-M., Kahn, S.\ M., Pisarski, R., \& Long, K.\ S.\ 1984, \apj, 278, 615 
\bibitem[Levenson \ea(1996)]{Lev96}Levenson, N. A., Graham, J. R., Hester, J. J., \& Petre, R. 1996, \apj, 468, 323 
\bibitem[Levenson \ea (1998)]{Lev98}Levenson, N. A., Graham, J. R., Keller, L. D., \& Richter, M. J. 1998, \apjs, 118, 541
\bibitem[Levenson \ea(1997)]{Lev97}Levenson, N. A., et al. 1997, \apj, 484, 304
\bibitem[Liedahl(1992)]{Lie92}Liedahl, D. A. 1992, Ph.D. Dissertation, University of California, Berkeley
\bibitem[Mewe, Gronenschild, \& van den Oord(1985)]{Mew85}Mewe, R., Gronenschild, E. H. B. M., \& van den Oord, G. H. J. 1985, \aaps, 62, 197
\bibitem[Mewe, Lemen, \& van den Oord(1986)]{Mew86}Mewe, R., Lemen, J. R., \& van den Oord, G. H. J. 1986, \aaps, 65, 511
\bibitem[Miyata \& Tsunemi(1999)]{Miy99}Miyata, E., \& Tsunemi, H.\ 1999, \apj, 525, 305
\bibitem[Miyata et al.(1994)]{Miy94} Miyata, E., Tsunemi, H., Pisarski, R., \& Kissel, S.~E.\ 1994, \pasj, 46, L101
\bibitem[Parker(1967)]{Par67} Parker, R. A. R. 1967, \apj, 149, 363 
\bibitem[Rasmussen \& Martin(1992)]{Ras92}Rasmussen, A.,\& Martin, C.\ 1992, \apjl, 396, L103. 
\bibitem[Raymond et al.(1980)]{Ray80}Raymond, J.~C., Davis, M., Gull, T.~R., \& Parker, R.~A.~R.\ 1980, \apjl, 238, L21 
\bibitem[Raymond et al.(1988)]{Ray88} Raymond, J.~C., Hester, J.~J., Cox, D., Blair, W.~P., Fesen, R.~A., \& Gull, T.~R.\ 1988, \apj, 324, 869
\bibitem[Scoville et al.(1977)]{Sco77}Scoville, N. Z., Irvine, W. M., 
Wannier, P. G., \& Predmore, C. R. 1977, \apj, 216, 320
\bibitem[Slavin, Shull, \& Begelman(1993)]{Sla93}Slavin, J.~D., Shull, J.~M., \& Begelman, M.~C.\ 1993, \apj, 407, 83 
\bibitem[Townsley et al.(2000)]{Tow00}Townsley, L.~K., {Broos}, P.~S., {Garmire}, G.~P., \& {Nousek}, J.~A. 2000, \apjl, 534, L139
\bibitem[Vedder et al.(1986)]{Ved86}Vedder, P.~W., Canizares, C.~R., Markert, T.~H., \& Pradhan, A.~K.\ 1986, \apj, 307, 269
\bibitem[Weisskopf et al.(2000)]{Wei00}Weisskopf, M.~C., Tananbaum, H.~D., Van Speybroeck, L.~P., \& O'Dell, S.~L.\ 2000, \procspie, 4012, 2 

\end{thebibliography}
